\documentclass[letterpaper]{aa}
\usepackage{epsfig,natbib}

\newcommand{\gcc}{\mbox{g cm$^{-3}$}}
\newcommand{\dync}{\mbox{dyn cm$^{-2}$}}
\newcommand{\req}[1]{Eq.~(\ref{#1})}
\newcommand{\dd}{\mathrm{d}}

\begin{document}

\title{Analytical representations of unified equations of state
 of neutron-star matter}

\author{P. Haensel\inst{1} \and A. Y. Potekhin\inst{2,3}}
 \institute{N. Copernicus Astronomical Center, Polish
       Academy of Sciences, Bartycka 18, PL-00-716 Warszawa, Poland
       \and
       Ioffe Physico-Technical Institute,
    Politekhnicheskaya 26, 194021 St.~Petersburg, Russia
    \and
    Isaac Newton Institute of Chile, St.~Petersburg Branch, Russia}

\offprints{P. Haensel,
\email{haensel@camk.edu.pl}}
\date{
Received  23 July 2004 / Accepted 12 August 2004
}
\abstract{
Analytical representations are derived for two
equations of state (EOSs) of neutron-star matter: FPS and SLy.
Each of these EOSs is unified,
that is, it describes the crust and the core of a neutron star
using the same
physical model. Two versions of the
EOS parametrization are considered. In the  first one,
pressure and mass density
are given as  functions of the baryon density. In the second version, pressure,
mass density,  and baryon density are given as functions of the pseudo-enthalpy,
which makes   this representation particularly useful for 2-D calculations
of stationary rotating configurations of neutron stars.
}
\titlerunning{Analytical EOS of neutron-star matter}
\authorrunning{P. Haensel \& A. Y. Potekhin}
\maketitle
\section{Introduction}
\label{sect:introd}
%
The equation of state (EOS) of dense matter is a crucial input for the
neutron-star structure calculations. Under standard conditions  neutron-star
matter is strongly degenerate, and therefore the matter pressure is
temperature independent; exceptions are
the outermost (a few meters thick) envelopes,
newly-born
neutron stars, and the envelopes of exploding X-ray bursters. At
$\rho\ga 10^8~\gcc$ the EOS is not affected by the magnetic
field even as strong as $10^{14}~$G and by the temperature $T\la 10^9$~K.
Therefore, except for a thin outer
envelope (which for the bottom density of $10^8~\mathrm{g~cm^{-3}}$ is
$\sim100$~m thick and contains only $\sim10^{-9}~\mathrm{M}_\odot$ of matter),
the EOS
of neutron-star matter has a one-parameter character. In order to determine
the neutron-star structure up to the maximum allowable mass, $M_\mathrm{max}$,
one has to know the EOS up to a few times $10^{15}~\mathrm{g~cm^{-3}}$. While the
EOS of the neutron-star crust ($\rho \la  (1-2)\times 10^{14}~\mathrm{g~cm^{-3}}$)
is rather well known (e.g., \citealt{Haens01}), the EOS of the liquid core at
$\rho\ga 5\times 10^{14}~\mathrm{g~cm^{-3}}$, which is crucial for determining
$M_\mathrm{max}$,  remains uncertain (e.g.,
\citealt{HeisPandh00,Haens03}).

The EOS of the crust depends on the crust formation scenario. Two limiting
cases are:  cold catalyzed matter being at the ground
state at a fixed baryon density $n$, and the accreted crust
formed via compression at $T<10^9~$K from the thermonuclear ashes of the
X-ray bursts in the outer envelopes of accreting neutron stars
\citep{HZ90,HZ03}. The EOS of the liquid core does not depend on the
formation scenario, and can slightly deviate from the ground state one
only when there are deviations from the weak interaction equilibrium 
-- e.g.,
when rapid matter flow (like in stellar pulsations) is involved.
In the present paper we consider the standard case of the
ground-state matter.

A ``unified EOS'' is obtained in the many-body calculations based on a
single effective nuclear Hamiltonian, and is valid in  all regions of the
neutron star interior. For unified EOSs the transitions between the outer
crust and the inner crust, and between the inner crust and the core are
obtained as a result of many-body calculation. Alas, up to now only a few
models of unified EOSs have been constructed.
All other EOSs consist of crust and core segments
obtained using different physical models. The crust-core interface has there
no physical meaning  and both segments are joined using an ad hoc matching
procedure. Therefore, neutron-star models based on these EOSs contain a
shell with an unphysical EOS. For such ``matched EOSs'' it is not
possible to study phenomena which are sensitive to the position of the
crust-core interface.

In the present paper we
consider two unified EOSs, the FPS EOS of \citet{Pandharipande89}
and the SLy EOS of \citet{DouchinHaensel01}.

EOSs are usually given in the form of tables. Therefore, in order to
use them,  one has to employ interpolation between the tabulated points.
However, the interpolation
procedure is not unique.  This introduces ambiguities in the calculated parameters
of the neutron star models. Moreover, interpolation should be
done respecting exact  thermodynamic  relations. This    turned out
to be a  particularly serious  problem in the high-precision
2-D calculations of models of rapidly rotating neutron stars
\citep{Nozawa98}. In the 3-D calculations of the stationary
configurations of a close binary neutron-star system one needs derivatives
of pressure with respect to the enthalpy, and  tabulated non-polytropic
EOSs are not useful in this respect (see, e.g., \citealt{GourgGrand01}).

In view  of the deficiencies and ambiguities inherent in the
 tabulated EOSs, which are particularly serious in the case of matched EOSs,
it is of great interest to derive analytical representations of the EOSs.
They  have two important advantages over the tabulated ones. First, there
is no ambiguity of interpolation and the derivatives can be
 precisely calculated. Second, these representations can be constructed
fulfilling exactly the thermodynamic relations. In this way, analytical EOSs
can allow for a very high precision of neutron star structure calculation
 in the 2-D and 3-D  cases.

In the present paper we derive analytical representations of the unified FPS
and SLy EOSs. In Sect.~\ref{sect:EOS.gen}
we discuss general properties of the EOS used
for the calculations of non-rotating and rotating  neutron-star
models. Analytical representations of the EOSs for non-rotating stars
are described in Sect.~\ref{sect:fits},
while Sect.~\ref{sect:Gamma} is devoted to the case
of rotating stars. 
Discussion and conclusion are presented in Sect.~\ref{sect:conclusion}.
\section{General properties of the one-parameter EOS of neutron-star matter}
\label{sect:EOS.gen}
A given EOS is usually presented in the form of a table containing a grid
of calculated values of matter density (full energy density ${\cal E}$
including
the rest energies of the matter constituents divided by $c^2$, i.e.,
$\rho = {\cal E}/c^2$), baryon (number) density $n$, and pressure $P$.
The EOS $\lbrace{\rho_i,n_i,P_i\rbrace}$ ($i=1,\ldots ,N$) is
then interpolated between the tabulated points so as to get the one-parameter
EOS in the form $P=P(n)$, $\rho=\rho(n)$. The  interpolation involves some
degree of indeterminacy (there are many ways of interpolating) and this itself
implies some ambiguity as far as the calculated values of the
neutron-star parameters
(for example, the value of $M_\mathrm{max}$) are concerned.

\subsection{Non-rotating configurations}
\label{sect:NS.nonrot}
High precision determination of the baryon number $A$ and gravitational mass
$M$ of an equilibrium configuration requires condition of thermodynamic
consistency of functions $\rho(n)$ and $P(n)$ to be strictly fulfilled. The
first law of thermodynamics
in the $T=0$ limit implies relation
\begin{equation}
P(n)=n^2 c^2 \frac{\mathrm{d}}{\mathrm{d} n}\left(\frac{\rho}{n}\right),
\label{eq:P.n.rho}
\end{equation}
which puts constraints on the interpolation procedure (see, e.g.,
\citealt{HaenseProsz82}); the above relation can be also used in the integral
forms,
\begin{eqnarray}
\frac{\rho(n)}{n} &=& \frac{\rho_\mathrm{s}}{n_\mathrm{s}} +
\int_{n_\mathrm{s}}^n {P(n^\prime)\over {n^\prime}^2 c^2}
\mathrm{d}n^\prime~,
\label{eq:rho.n.Integral}
\\
\ln\left(\frac{n}{n_\mathrm{s}}\right) &=& c^2
 \int_{\rho_\mathrm{s}}^\rho
 \frac{\mathrm{d}\rho'}{P(\rho') + \rho' c^2} ~,
\label{eq:n.Integral}
\end{eqnarray}
where $\rho_\mathrm{s}$ and $n_\mathrm{s}$ are the values of $\rho$ and $n$ at
the neutron-star surface.
In the present paper we put $\rho_{\rm s}$ equal to density of $^{56}{\rm Fe}$
at zero pressure and zero temperature,
$\rho_{\rm s}=7.86~{\rm g~cm^{-3}}$. In the outermost neutron-star layers,
we fix the value
of mass per nucleon as $m_0=1.66\times 10^{-24}~$g, so that $n_{\rm s}=
\rho_{\rm s}/m_0=4.73494\times 10^{24}~{\rm cm^{-3}}$.

An example of an interpolation recipe which respects 
Eqs.\ (\ref{eq:P.n.rho})--(\ref{eq:n.Integral}) and at the same
time yields highly smooth functions $P(n)$ and $\rho(n)$ was
presented by \citet{Swesty96}. Only a thermodynamically consistent
interpolation yields neutron-star models which strictly satisfy
the relation connecting the baryon chemical potential
$\mu_\mathrm{b}=c^2 {\rm d}\rho/{\rm d}n_{\rm b}=
(\rho c^2 +P)/n$ and the metric function $\Phi(r)$ at
a given circumferential radius $r$.
This strict relation stems from the equation for the
metric function
\begin{equation}
{{\rm d}\Phi\over {\rm d}r}=-{\rho c^2\over \rho c^2 +P}{{\rm
d}P\over {\rm d}r}~,
\label{eq:Phi.TOV}
\end{equation}
and can be written as
\begin{equation}
\mu_\mathrm{b}(r) \mathrm{e}^{\Phi(r)} =
         \mu_\mathrm{b}(R)
\mathrm{e}^{\Phi(R)}~,
\label{eq:mu_b.const}
\end{equation}
where $\Phi(R)=\sqrt{1-2GM/Rc^2}$. Here, $M$ and $R$ are
the total gravitational mass and circumferential
stellar radius, respectively.
Strictly speaking, if Eq.~(\ref{eq:mu_b.const})
does not hold, the calculated configuration
{\it is not} in hydrostatic equilibrium. Alas, this often happens
for the neutron-star models calculated using tabulated EOSs with
logarithmic interpolation between the tabulated points (see, e.g.,
\citealt{Harrison65,BPS,ArnettBowers77}).
Resulting inconsistencies may seem minor, but they may lead  to
serious problems if high precision of a {\it simultaneous}
determination of $M$ and $A$ is concerned; this is the case of the
energy release due to a phase transition in the neutron-star core (see,
e.g., \citealt{HaenseProsz82}). On the contrary, if Eq.\
(\ref{eq:mu_b.const}) is strictly fulfilled, then the constancy of
$\mu_\mathrm{b}\mathrm{e}^\Phi$ and the accuracy of calculating
stellar parameters are limited only by the numerical precision of
the computer code.

For a thermodynamically consistent EOS, Eq.\ (\ref{eq:mu_b.const})
implies the baryon density profile within a static neutron star,
\begin{equation}
n(r)=n_{\rm s}\left[{\rho(r)\over \rho_{\rm s}} + {P(r)\over
\rho_{\rm s}c^2}\right]{\mathrm e}^{\Phi(r)-\Phi(R)}~,
\label{eq:nb.prof.stat}
\end{equation}
where we have neglected $P_\mathrm{s}=P(R,T)$ because
the pressure within the atmosphere
is small compared with $\rho_\mathrm{s} c^2$.
Therefore, what  one needs to get $n(r)$ in a static neutron star is
just the surface baryon density, $n_{\rm s}$; the $\lbrace{ n_i \rbrace}$
column of the tabulated EOS turns out to be  redundant.

\subsection{Stationary rotating configurations}
\label{sect:NS.rot}
Rotation breaks the spherical symmetry of the equilibrium configuration.
Stationary configurations of a rigidly rotating star
of ideal liquid are solutions of 2-D axially symmetric partial differential
equations of
hydrostatic equilibrium in coordinates $r$ and $\theta$ (rotation of
relativistic stars is reviewed by \citealt{Sterg03}). As far as the EOS is
concerned, it is suitable to parametrize it in terms of a dimensionless
{\it pseudo-enthalpy}
\begin{equation}
H(P)\equiv
\int_0^P{\mathrm{d}P^\prime\over \rho (P^\prime)c^2 +P^\prime}~.
\label{eq:H.def}
\end{equation}
It  can be rewritten in terms of the enthalpy per baryon
\begin{equation}
h(P)={\rho c^2 +P\over n}
\label{eq:h.def}
\end{equation}
as
\begin{equation}
H(P)=\mathrm{ln}\left[{h(P)\over h_{\rm s}}\right]~,
\label{eq:H.h}
\end{equation}
where $h_{\rm s}=\rho_{\rm s}c^2/n_{\rm s}$.
The pseudo-enthalpy $H$ vanishes at the stellar surface and increases
monotonically  towards the stellar center. It is a very useful variable
in numerical calculations of the stationary configurations of rotating
stars, because the definition of $H$ allows one
to easily write down the
first integral of motion
\begin{equation}
H(r,\theta)+\Phi(r,\theta)-\ln\gamma(r,\theta)=
H(0,0)+\Phi(0,0)~,
\label{eq:H.1st.integral}
\end{equation}
where $\Phi$ is the metric function and $\gamma$ is the Lorentz factor
connecting the two  measured values of a physical quantity: that measured
in a local observer's reference frame at a point $(r,\theta,\phi)$
and that measured by the distant observer in an  inertial reference frame at
infinity,
\begin{equation}
\gamma = \left( 1- U^2\right)^{-1/2}~,
\label{eq:Gamma.U}
\end{equation}
where $U$ is the fluid velocity in the azimuthal ($\phi$) direction,
as measured by a
local observer. Equation (\ref{eq:H.1st.integral}) is the general relativistic
version of the Bernoulli theorem for a stationary ideal fluid flow in
gravitational field. Consequently, the most useful parametrization of the
EOS for rotating stars is $\rho=\rho(H)$, $P=P(H)$. Then, the baryon number
density is given by an exact analytical formula
\begin{equation}
n(H)=
n_{\rm s}\left[{\rho(H)\over \rho_{\rm s}}+
{P(H)\over \rho_{\rm s} c^2}\right]
{\mathrm{e}^{-H}} ~,
\label{eq:n_b.prof.rot}
\end{equation}
which is the rotating-star analogue  of  Eq.\ (\ref{eq:nb.prof.stat}).
\section{Analytical representations of the EOS}
\label{sect:fits}
There are three qualitatively different domains of the interior of a
neutron star, separated by phase transition points:
the outer crust (consisting of the electrons and atomic nuclei),
the inner
crust (consisting of the electrons, nuclei, and dripped neutrons), and
the core
which contains the electrons, neutrons, protons, $\mu^-$-mesons,
and possibly $\pi$- and $K$-mesons, some hyperons, or
quark matter. The latter species are contained in the innermost stellar
domain called the inner core. In addition, there can be density
discontinuities at the interfaces between layers containing different
nuclei in the crust.
In the fitting, we neglect these small
discontinuities and approximate the EOS by fully analytical functions.
However, the different character of the EOS in the different domains
is reflected by the complexity of the fit, which consists of
several fractional-polynomial parts, matched together by virtue of the
function
\begin{equation}
    f_0(x) = \frac{1}{\mathrm{e}^x+1}~.
\end{equation}

We rely on a tabulated unified EOS (FPS\footnote{The FPS
table has been kindly provided by N.\ Stergioulas.}
or SLy) at
 $\rho>5\times10^{10}$ \gcc. At lower densities,
$10^8~\gcc\la\rho<5\times10^{10}$ \gcc, the crustal matter is described
by the EOS of \citet{HP94} (HP94), based on experimental nuclear data,
supplemented by the EOS for cold catalyzed matter due to
\citet{BPS} (BPS)
at still lower density $\rho\la10^8$ \gcc.
The lowest-density parts of the tables at $\rho < 10^5~\gcc$ have not
been used in the fitting, because at such low density the EOS is no
longer one-parametric, but depends also on temperature (see
Fig.~\ref{fig:fit_sly}).

\begin{table}[!t]
\centering
\caption[]{Parameters of the fit (\protect\ref{eq:fit.P})}
\label{tab:fit.P}
\begin{tabular}{rll|rll}
\hline\hline\rule[-1.4ex]{0pt}{4.3ex}
i & $a_i$(FPS) & $a_i$(SLy) & i & $a_i$(FPS) & $a_i$(SLy)  \\
\hline\rule{0pt}{2.7ex}
$1$    & {6.22} &{6.22}        & ${10}$ &~\, 11.8421   &~\, 11.4950\\
$2$    & {6.121} & {6.121}     & ${11}$ &   $-22.003$  & $-22.775$ \\
$3$    & 0.006004 & 0.005925   & ${12}$ &~\, 1.5552   &~\, 1.5707 \\
$4$    & 0.16345   & 0.16326   & ${13}$ &~\, 9.3      &~\, 4.3    \\
$5$    & 6.50      & 6.48      & ${14}$ &~\, 14.19    &~\, 14.08  \\
$6$    & 11.8440   & 11.4971   & ${15}$ &~\, 23.73    &~\, 27.80  \\
$7$    & 17.24     & 19.105    & ${16}$ &  $-1.508$   & $-1.653$  \\
$8$    & 1.065     & 0.8938    & ${17}$ &~\, 1.79     &~\, 1.50   \\
$9$    & 6.54      & 6.54      & ${18}$ &~\, 15.13    &~\, 14.67 \rule[-1.4ex]{0pt}{0pt}\\
\hline\hline
\end{tabular}
\end{table}

\subsection{Non-rotating stars}
\label{sect:Analyt.EOS.static}
For non-rotating configurations,
we have  parametrized the pressure as
function of density.
Let us denote $\xi=\log(\rho/\textrm{g cm}^{-3})$,
 $\zeta = \log(P/\dync)$.
Then the  parametrization reads
\begin{eqnarray}
  \zeta &=& \frac{a_1+a_2\xi+a_3\xi^3}{1+a_4\,\xi}\,f_0(a_5(\xi-a_6))
\nonumber\\&&
     + (a_7+a_8\xi)\,f_0(a_9(a_{10}-\xi))
\nonumber\\&&
     + (a_{11}+a_{12}\xi)\,f_0(a_{13}(a_{14}-\xi))
\nonumber\\&&
     + (a_{15}+a_{16}\xi)\,f_0(a_{17}(a_{18}-\xi)) ~.
\label{eq:fit.P}
\end{eqnarray}
The parameters $a_i$ for FPS and SLy EOSs are given in
Table~\ref{tab:fit.P}. The typical fit error of $P$ is
1--2\% (for $\xi\ga5$). The maximum error is determined
by the jumps near the phase transitions in the tabulated EOSs,
which are smoothed by the fit (\ref{eq:fit.P}).
For FPS, the maximum error is 3.6\% at $\xi=14.22$
(crust-core interface).
For SLy, the maximum error is 2.9\% at $\xi=8.42$
($^{62}$Ni-$^{64}$Ni interface in the HP94 part of the table).

\begin{figure}[t!]
\centering
\epsfxsize=86mm
\epsffile{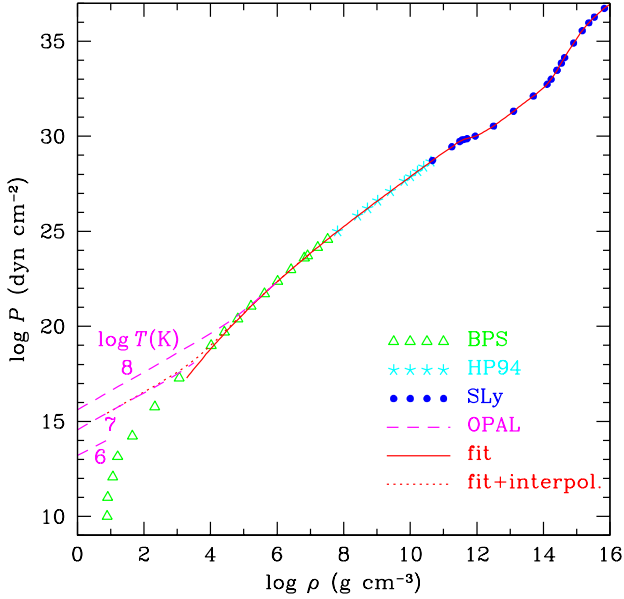}
\caption{Neutron-star EOS for non-rotational configurations:
BPS (triangles), \protect\citet{HP94} (HP94, stars),
SLy (dots), OPAL at $T=10^6$, $10^7$, and $10^8$~K
(dashed lines), the fit (\ref{eq:fit.P}) (solid line) and the fit modified
at low $\rho$ (dotted line).}
\label{fig:fit_sly}
\end{figure}

Figure \ref{fig:fit_sly} shows $\log P$ against $\log\rho$
for a tabulated EOS (symbols) and the 
corresponding fit
(solid line). Triangles correspond to BPS, stars to HP94, and
dots to SLy data. By construction, the fit is accurate at
$\rho\ga10^5$ \gcc. As stated above, at lower density the EOS becomes
temperature-dependent (for $T$ values typical for neutron star
envelopes). This is illustrated by the dashed lines, that show the
OPAL  EOS \citep{OPAL-EOS} for $T=10^6$, $10^7$, and
$10^8$~K.\footnote{ The OPAL table for iron has been kindly provided by
F.~J.\ Rogers.} However, a reasonable continuation of the fit to lower
densities can be  constructed by a simple interpolation. For instance,
the dotted line in  Fig.~\ref{fig:fit_sly} corresponds to
$P=P_\mathrm{fit}+P_0$, where $P_\mathrm{fit}=10^\zeta$ is given  by
\req{eq:fit.P} (where we always assume $\xi>0$), 
and $P_0=3.5\times10^{14}\,\rho$ approximates the OPAL
EOS near $\rho\sim\rho_\mathrm{s}$ at $T=10^7$~K
(here $P$ is in \dync\ and $\rho$ in \gcc).

\begin{figure}[t!]
\centering
\epsfxsize=86mm
\epsffile{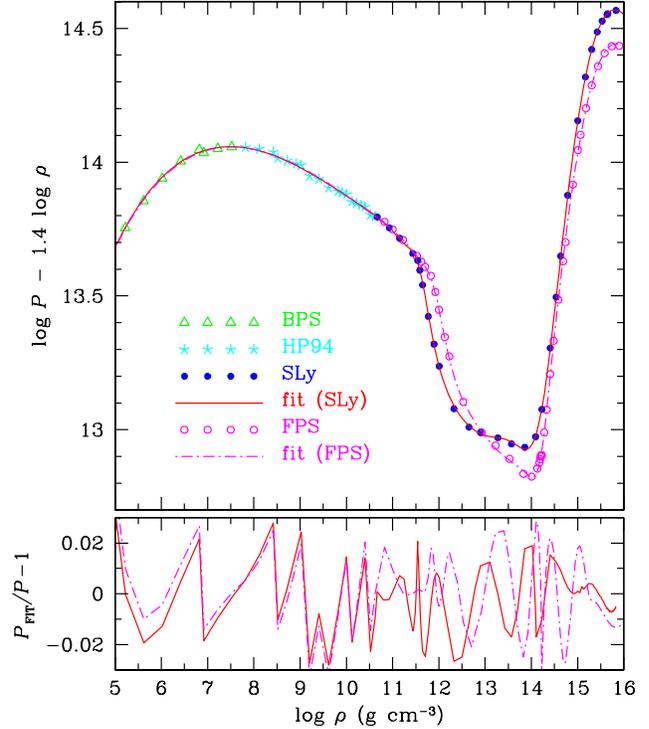}
\caption{Comparison of the data and fits for SLy and FPS EOSs
for non-rotational configurations. Upper panel:
rarefied tabular data (symbols) and the fit (\ref{eq:fit.P}) (lines);
lower panel: relative difference between the data and fit.
Filled dots and solid
line: SLy; open circles and dot-dashed line: FPS (triangles and stars
on the upper panel
are BPS and HP94 data at $\rho<5\times10^{10}$ \gcc).}
\label{fig:fit_nr}
\end{figure}

\begin{table}[!t]
\centering
\caption[]{Parameters of the fits (\protect\ref{eq:fit.rho-n})
and (\protect\ref{eq:fit.n-rho})}
\label{tab:fit.n.rho}
\begin{tabular}{r|ll|ll}
\hline\hline\rule[-1.4ex]{0pt}{4.3ex}
$i$ & $p_i$(FPS) & $p_i$(SLy) & $q_i$(FPS) & $q_i$(SLy) \\
\hline\rule{0pt}{2.7ex}
$1$ & 0.320 & 0.423 & 0.608 & 0.183 \\
$2$ & 2.17  & 2.42  & 2.41  & 1.26  \\
$3$ & 0.173 & 0.031 & 2.39  & 6.88  \\
$4$ & 3.01  & 0.78  & 3.581 & 3.612 \\
$5$ & 0.540 & 0.238 & 1.681 & 2.248 \\
$6$ & 0.847 & 0.912 & 0.850 & 0.911 \\
$7$ & 3.581 & 3.674 & 11.64 & 11.56 \rule[-1.4ex]{0pt}{0pt}\\
\hline\hline
\end{tabular}
\end{table}

In Fig.~\ref{fig:fit_nr} we compare the FPS and SLy EOSs. Symbols 
on the upper panel show
the data (triangles, stars, dots, and open circles for BPS, HP94,
SLy, and FPS, respectively) and lines show the fit (solid for SLy and
dot-dashed for FPS). In order to make the differences between the data
and fits and between SLy and FPS EOSs visible, we plot  the
\emph{difference} $\log P - 1.4 \log \rho$, where $P$ is in \dync\ and
$\rho$ in \gcc.
The lower panel shows the relative difference between the tabulated
and fitted EOSs (solid and dot-dashed lines for SLy and FPS, respectively).
It illustrates the accuracy of the fit (\ref{eq:fit.P}).

Now, $n(\rho)$ can be easily obtained from Eq.\ (\ref{eq:n.Integral}).
Doing this and substituting $P(\rho')$ in the integrand of
\req{eq:n.Integral} from \req{eq:fit.P},
we recover the original tabular values
with maximum difference $<0.4$\%
for FPS and $<0.12$\% for SLy.

\begin{figure}[t!]
\centering
\epsfxsize=86mm
\epsffile{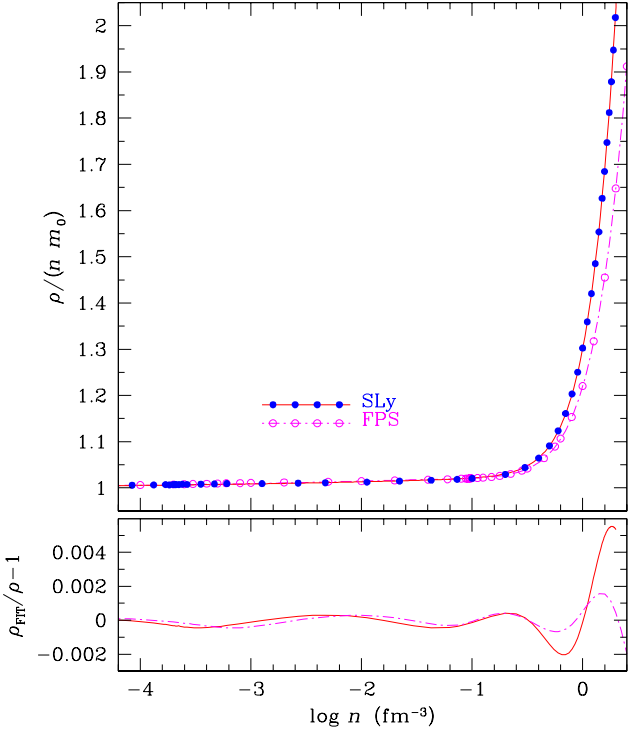}
\caption{Dependence of $\rho$ on $n$.
Upper panel: rarefied data (symbols)
 and the fit (\ref{eq:fit.rho-n}) (lines);
lower panel: relative difference between the data and fit.
Filled dots and solid
line: SLy; open circles and dot-dashed line: FPS. }
\label{fig:fit.n.rho}
\end{figure}

In some applications, it may be convenient to use $n$ as independent
variable, and treat $\rho$ and $P$ as functions of $n$. For this purpose
one can use the following fit:
\begin{eqnarray}
   \frac{\rho}{n m_0} &=& 1+
     \frac{p_1 n^{p_2}+p_3 n^{p_4}}{(1+p_5 n)^2}\,
     f_0(-{p_6}(\log n + p_7))
\nonumber\\&&
     +  
    \frac{n}{8\times10^{-6} + 2.1\,n^{0.585}}\,
     f_0({p_6}(\log n + p_7))
     ~,
\label{eq:fit.rho-n}
\end{eqnarray}
where $n$ is in fm$^{-3}$.
The inverse fit $n(\rho)$ is given by the formula
\begin{eqnarray}
   \frac{x}{n} &=& 1+
     \frac{q_1 x^{q_2}+q_3 x^{q_4}}{(1+q_5 x)^3}\,
     f_0({q_6}(q_7 - \log \rho))
\nonumber\\&&
     +  
    \frac{x}{8\times10^{-6} + 2.1\,x^{0.585}}\,
     f_0({q_6}(\log \rho - q_7)),
\label{eq:fit.n-rho}
\end{eqnarray}
where $x=\rho/m_0$ and $\rho$ is in \gcc.
Coefficients $p_i$ and $q_i$ of the fits
(\ref{eq:fit.rho-n}) and (\ref{eq:fit.n-rho}) are given in
Table~\ref{tab:fit.n.rho}.
The fractional fit residual in $\rho$ and $n$ varies from 
$\la10^{-9}$ at $\rho\la10$ {\gcc}
to a fraction of percent near the maximum $\rho\sim10^{16}$ \gcc,
while the difference $(\rho-n m_0)$
is approximated with the typical accuracy of a few percent.
In Fig.~\ref{fig:fit.n.rho} (upper panel) 
we plot the ratio $\rho/(n m_0)$
against $n$ for the two EOSs.
We see that in the crust, at $n\la0.1$ fm$^{-3}$,
$\rho$ is to a good accuracy proportional to $n$.
The lower panel, which shows the relative difference between 
fitted and tabulated values of $\rho(n)$, confirms the accuracy
of \req{eq:fit.rho-n}.

It should be stressed that thermodynamics requires \req{eq:P.n.rho} to
be satisfied exactly. To achieve this, one should not totally rely on
the fits (\ref{eq:fit.rho-n}) and (\ref{eq:fit.n-rho}); otherwise
thermodynamic consistency will be violated on the scale of these fits'
errors (a fraction of percent). So, if $\rho$ is used as an input, then
$n_\mathrm{b}(\rho)$ should be calculated from \req{eq:n.Integral}.
Alternatively, if the input is $n_\mathrm{b}$, then, having calculated
$\rho_\mathrm{fit}(n_\mathrm{b})$ from \req{eq:fit.rho-n} and
$P(n_\mathrm{b})=P(\rho_\mathrm{fit}(n_\mathrm{b}))$ from
\req{eq:fit.P}, one should refine $\rho(n_\mathrm{b})$ using relation
(\ref{eq:rho.n.Integral}).

\subsection{Rotating stars}
\label{sect:Analyt.EOS.rot}
\begin{figure}[t!]
\centering
\epsfxsize=86mm
\epsffile{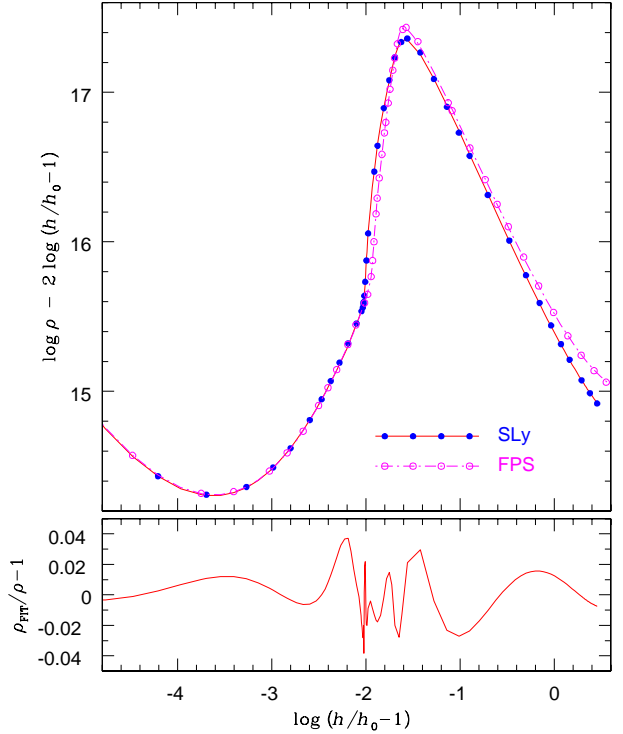}
\caption{SLy and FPS EOS for rotational configurations.
Upper panel: rarefied data to be fitted, calculated according to
Eq.~(\ref{eq:fit.P}) (symbols)
 and the fit (\ref{eq:fit.rho}) (lines);
 lower panel: relative difference between the data and fit.
Filled dots and solid
line: SLy; open circles and dot-dashed line: FPS. }
\label{fig:hfit}
\end{figure}

As explained in Sect.~\ref{sect:NS.rot}, for rotating stars
it is most useful to parametrize
density and pressure as functions of the pseudo-enthalpy $H$,
which can be written in terms of the enthalpy per baryon $h$.
Let us define $\eta\equiv h/m_0 c^2-1$.
In view of relation (\ref{eq:h.def}), the function which we
intend to parametrize, $\xi(\eta)$,
 is not independent of the function $\zeta(\xi)$
parametrized by \req{eq:fit.P}.
 In order to fulfill Eq.~(\ref{eq:h.def}) as closely as possible,
 we first calculate $\eta(\xi)$
  using Eqs.~(\ref{eq:fit.P}) and (\ref{eq:h.def}),
and then find the inverse fit $\xi(\eta)$.
The best fit reads:
\begin{eqnarray}
  \xi &=& \left[ b_1+b_2 \log\eta+\frac{b_3\eta^{b_4}}{1+b_5\eta}
       \right] f_0(b_6(\log\eta-b_7))
\nonumber\\&&
     +  \frac{b_8+b_9\log\eta+(b_{10}+b_{11}\log\eta)(b_{12}\eta)^7}{
        1+b_{13}\eta+(b_{12}\eta)^7}
\nonumber\\&&\times f_0(b_6(b_7-\log\eta))
     +  b_{14}\,f_0(b_{15}(b_{16}-\log\eta))~.
\label{eq:fit.rho}
\end{eqnarray}
where the parameters $b_i$ are given in Table~\ref{tab:fit.rho}.
The comparison of the fit and the data is presented 
in Fig.~\ref{fig:hfit}.
The typical fit error of $\rho$, according to Eq.~(\ref{eq:fit.rho}),
is about 1\% at $\eta\ga10^{-7}$
(corresponding to $\xi\ga 3$),
and the maximum error $< 4$\% occurs near the neutron drip
and crust-core phase transitions.

\begin{table}[!t]
\centering
\caption[]{Parameters of the fit (\protect\ref{eq:fit.rho})}
\label{tab:fit.rho}
\begin{tabular}{rll|rll}
\hline\hline\rule[-1.4ex]{0pt}{4.3ex}
$i$ &~\, $b_i$(FPS) &~\, $b_i$(SLy) & $i$ & $b_i$(FPS) & $b_i$(SLy) \\
\hline\rule{0pt}{2.7ex}
$1$    &~\, 5.926  &~\, 5.926   &$9$    & 11.97  & 34.96  \\
$2$    &~\, 0.4704 &~\, 0.4704  &${10}$ & 15.432 & 15.328 \\
$3$    &~\, 19.92  &~\, 20.13   &${11}$ & 0.6731 & 0.621  \\
$4$    &~\, 0.2333 &~\, 0.2347  &${12}$ & 49.4   & 63.1   \\
$5$    &~\,  2.63  &~\, 3.07    &${13}$ & 11.47  & 68.5   \\
$6$    &~\, 54.7   &~\, 97.8    &${14}$ & 1.425  & 2.518  \\
$7$    & $-1.926$  &  $-2.012$&${15}$   & 3.0    &  2.6   \\
$8$    &~\, 36.89  &~\, 89.85  &${16}$  & 0.913  & 1.363 \rule[-1.4ex]{0pt}{0pt} \\
\hline\hline
\end{tabular}
\end{table}

When used in combination, the fits (\ref{eq:fit.P}) 
and (\ref{eq:fit.rho}),
together with Eq.~(\ref{eq:n.Integral}) or \req{eq:fit.n-rho}
give the parametrizations of $\rho(H)$,
$P(H)$, and $n(H)$ needed for calculations of
the stationary rotating configurations.
In this case, the function $P(H)=P(\rho(H))$ obtained using 
Eqs.~(\ref{eq:fit.rho}) and (\ref{eq:fit.P}), reproduces
the tabular values with a typical discrepancy of 1--2\%,
with a maximum within 10\% near the crust-core boundary.

The remark on the thermodynamic consistency, made at the end 
of Sect.~\ref{sect:Analyt.EOS.static}, applies also here:
one should refine either $n_\mathrm{b}$ or $\rho$ 
fitted values, using the exact relations
(\ref{eq:n.Integral}) or (\ref{eq:rho.n.Integral}).

\section{Adiabatic index}
\label{sect:Gamma}
\begin{figure}[t!]
\centering
\epsfxsize=86mm
\epsffile{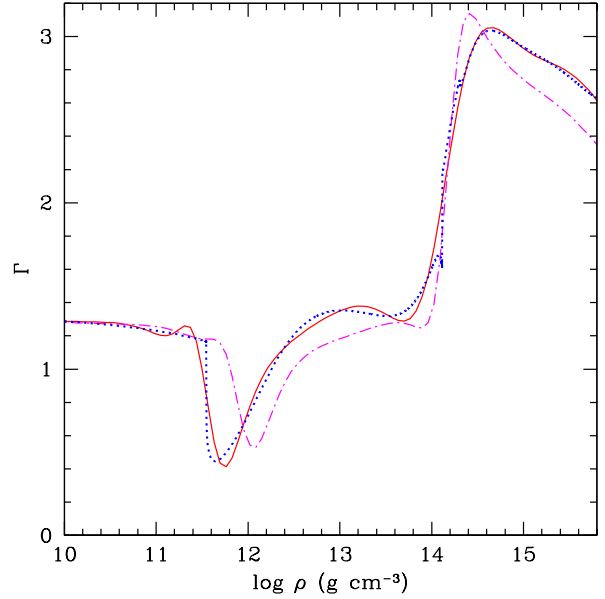}
\caption{Adiabatic index $\Gamma$ for SLy and FPS EOSs. 
Solid line: analytical approximation (SLy);
dotted line: precise values (SLy);
dot-dashed line: analytical approximation (FPS).}
\label{fig:Gamma}
\end{figure}

An important dimensionless parameter characterizing the stiffness of
the EOS at given density is the adiabatic index, defined by
\begin{equation}
   \Gamma=\frac{n}{P}\,\frac{\dd P}{\dd n}
   = \left[ 1+ \frac{P}{\rho c^2} \right]
   \frac{\rho}{P}\frac{\dd P}{\dd\rho}
~.
\end{equation}
Using our fit (\ref{eq:fit.P}), we obtain
 the analytical expression
\begin{eqnarray}&&\!\!\!\!\!
   \frac{\rho}{P}\frac{\dd P}{\dd\rho} =
   \frac{\dd\zeta}{\dd\xi}
   = \bigg[ \frac{
     a_2-a_1 a_4 + 3a_3\xi^2+2a_3 a_4 \xi^3}{(1+a_4\xi)^2}
\nonumber\\&&\hspace*{1em}
    -a_5\,\frac{1_1+a_2\xi+a_3\xi^3}{1+a_4\xi}\,f_0(a_5(a_6-\xi))\bigg]
    f_0(a_5(\xi-a_6))
\nonumber\\&&
   + \sum_{i=2}^4 f_0(a_{4i+1}(a_{4i+2}-\xi))
          \big[ a_{4i}
\nonumber\\&&\hspace*{1em}
      +a_{4i+1} (a_{4i-1}+a_{4i}\xi)f_0(a_{4i+1}(\xi-a_{4i+2})) \big]~.
\label{eq:fit.Gamma}
\end{eqnarray}
Different regions of neutron-star interior are characterized by distinct
behavior of $\Gamma$, displayed in Fig.~\ref{fig:Gamma}. Precise values
of $\Gamma$ calculated by \citet{DouchinHaensel01} are shown by dotted line,
and the analytical approximations according to Eqs.~(\ref{eq:fit.P}),
(\ref{eq:fit.Gamma}) by solid line (for SLy) and dot-dashed line (for FPS).

In the outer crust, the value
of $\Gamma$ depends quite weakly on density.
At $10^8~\gcc\la\rho\la3\times10^{11}~\gcc$,
this value would be $\Gamma\approx4/3$ if $A,Z$ values were fixed,
because in this case $P$ is mainly determined by
the pressure of ultrarelativistic electron gas
which behaves as $\propto(Z\rho/A)^{4/3}$.
For example, $\Gamma\approx4/3$ within each shell with constant $A$ and $Z$ of
the EOS derived by \citet{HP94}.
However, the compressible liquid drop model used by \citet{DouchinHaensel01}
effectively smoothes the discontinuities caused by transitions
from one to another ($A,Z$)-species with increasing density,
which leads to an effective continuous increase of the $A/Z$ ratio
and corresponding decrease of $\Gamma$, seen in Fig.~\ref{fig:Gamma}.

A dramatic drop in $\Gamma$ occurs at neutron drip threshold,
which corresponds to strong softening of the EOS.
The analytical expression (\ref{eq:fit.Gamma})
 somewhat smoothes this drop.
The behavior of
$\Gamma$ in the inner crust results from an interplay of several
factors, with stiffening due to interaction
between dripped neutrons, a softening effect of
neutron-gas -- nuclear-matter coexistence, and
the softening Coulomb contribution.

At the crust-core interface, matter strongly stiffens, and $\Gamma$
jumps from $\approx1.7$ to $\approx2.2$, which results from disappearance of
nuclei. The analytical approximation smoothes this jump also,
though reflects the stiffening. This approximation is also smooth
across a small discontinuous drop of $\Gamma$
at $\rho\approx2\times10^{14}~\gcc$,
where muons start to replace a part of ultrarelativistic electrons.
However, the electrons and muons give only minor
contribution to the pressure (and therefore behavior of $\Gamma$)
in the core, because the main contribution comes from
interactions between nucleons.
\section{Summary, discussion,  and conclusion}
\label{sect:conclusion}
 Analytical representations of the EOSs
used in the modern 2-D and 3-D simulations of neutron star
dynamics have many important advantages over the tabulated EOSs.
Analytical EOSs do not require any interpolations, are
thermodynamically consistent, and allow for a very high precision
of calculations. In the present paper we constructed analytical
representations, in terms of the continuous and differentiable
functions of a single chosen variable,  of the SLy and FPS EOSs.
For these analytical representations, the thermodynamic relations
are exactly satisfied at any point. Two choices of the independent
variable were considered. The first one is $\rho$; function
$P(\rho)$ given by \req{eq:fit.P}
fits the original tables in the density
interval $10^{5}~{\rm g~cm^{-3}}<\rho<10^{16}~{\rm g~cm^{-3}}$
within typical error of 1--2\%. 
Function $n(\rho)$ can be calculated either from \req{eq:n.Integral}
to satisfy exactly the first law of thermodynamics, 
or from the fit (\ref{eq:fit.n-rho}) with a typical error $\sim0.1$\%.
A variant which ensures the same accuracy is to choose $n$
as an independent variable and calculate $\rho(n)$ 
from the fit (\ref{eq:fit.rho-n}) and $P(\rho)$ from \req{eq:fit.P}.
The other choice of the independent
variable is to use the pseudo-enthalpy $H$. This choice is
particularly advantageous for applications to 2-D and 3-D
numerical simulations of neutron star dynamics, such as rotation
and inspiraling stage of the evolution of relativistic
neutron-star -- neutron-star binary. We represented both EOSs by
the continuous and differentiable functions $P(H)$, $\rho(H)$, and
$n(H)$, where $\rho(H)$ is given by \req{eq:fit.rho}
with typical accuracy within a few percent, while $P(H)$ and $n(H)$
are calculated from the functions $P(\rho)$ and $n(\rho)$, respectively.
Differentiation of $P(\rho)$ then yields
analytical representations of the
adiabatic index for the SLy and FPS EOSs; 
this quantity is important, for example, for
numerical simulations of dynamics of the neutron-star --
neutron-star system at the inspiral phase.

The quality of our  analytical representations of the EOSs
was tested by evaluation of the virial identities GRV2
\citep{Bona73,BonaGourg94}
and GRV3 \citep{GourgBona94}
in the numerical simulations of the 2-D stationary rotation 
of neutron stars. GRV2 and GRV3 are integral identities which 
must be satisfied by a stationary solution of 
the Einstein equations 
 and which \emph{are not imposed} in the numerical procedure
(see \citealt{Nozawa98} for the details of computation of GRV2 and GRV3).
For rotating configurations we get 
$\mathrm{GRV2,\,GRV3}\sim 10^{-6}-10^{-5}$, which is excellent.  

The subroutines for the numerical applications of analytical EOSs
(Fortran and C++ versions) can be
downloaded from the public domain 
\texttt{http://www.ioffe.ru/astro/NSG/NSEOS/}.
\begin{acknowledgements}
We are very grateful to M.~Bejger for performing the 2-D 
calculations for rapidly rotating neutron stars 
with our EOSs, and to M.~Shibata for pointing out to 
significant misprints in Eqs.~(\ref{eq:fit.rho-n})
and (\ref{eq:fit.n-rho}) of the preliminary version of this paper.
The work of P.H.\ was partially supported by the KBN grant 
no.~1-P03D-008-27.
The work of A.P.\ was partially supported by
the RFBR grants 02-02-17668 and 03-07-90200
and the Russian Leading Scientific Schools grant 1115.2003.2. 
\end{acknowledgements}


\begin{thebibliography}{}
\bibitem[Arnett \& Bowers(1977)]{ArnettBowers77}
Arnett, W.~D., \& Bowers, R.~L. 1977,
ApJS, 33, 415
\bibitem[Baym et al.(1971)Baym, Pethick, \& Sutherland]{BPS}
Baym, G., Pethick, C., \& Sutherland, P., 1971,
ApJ, 170, 299
\bibitem[Bonazzola(1973)]{Bona73}
Bonazzola S., 1973, ApJ, 182, 335
\bibitem[Bonazzola \& Gourgoulhon(1994)]{BonaGourg94}
Bonazzola S., Gourgoulhon E., 1994, Class.\ Quantum Grav., 
11, 1775
\bibitem[Douchin \& Haensel(2001)]{DouchinHaensel01}
Douchin, F., \& Haensel, P., 2001, A\&A, 380, 151
\bibitem[Gourgoulhon \& Bonazzola(1994)]{GourgBona94}
Gourgoulhon, E., \& Bonazzola S. 1994, Class.\ Quantum Grav., 
11, 443
\bibitem[Gourgoulhon et al.(2001)]{GourgGrand01}
Gourgoulhon, E., Grandcl{\'e}ment, P., Taniguchi, K., Marck, J.-A.,
\& Bonazzola, S., 2001, Phys.\ Rev. D, 63, 064029
\bibitem[Haensel(2001)]{Haens01}
Haensel, P., 2001,
Lecture Notes in Physics, 578, 127
\bibitem[Haensel(2003)]{Haens03}
Haensel, P., 2003,
Equation of State of Dense Matter and Maximum Mass of Neutron Stars,
in Final Stages of Stellar Evolution, ed. C. Motch
 \& J.-M. Hameury, EAS Publications Ser., vol.~7 (Les Ulis: EDP), 249
\bibitem[Haensel \& Pichon(1994)]{HP94}
Haensel, P., \&  Pichon, B., 1994, A\&A, 283, 313
\bibitem[Haensel \& Proszynski(1982)]{HaenseProsz82}
Haensel, P., \&  Proszynski, M., 1982, ApJ, 258, 306
\bibitem[Haensel \& Zdunik(1990)]{HZ90}
Haensel, P., \& Zdunik, J.~L. 1990,
A\&A, 229, 117
\bibitem[Haensel \& Zdunik(2003)]{HZ03}
Haensel, P., \& Zdunik, J.~L. 2003,
A\&A, 404, L33
\bibitem[Harrison et al.(1965)]{Harrison65}
Harrison, B.~K., Thorne, K.~S., Wakano, M., \& Wheeler, J.~A.
1965, Gravitation Theory and Gravitational Collapse (Chicago:
University of Chicago Press)
\bibitem[Heiselberg \& Pandharipande(2000)]{HeisPandh00}
Heiselberg, H., \& Pandharipande, V.~R. 2000,
Ann.\ Rev.\ Nucl.\ Part.\ Sci., 50, 481
\bibitem[Nozawa et al.(1998)]{Nozawa98}
Nozawa, T., Stergioulas, N., Gourgoulhon, E., \& Eriguchi, Y. 1998,
A\&AS, 132, 431
\bibitem[Rogers et al.(1996)Rogers, Swenson, \& Iglesias]{OPAL-EOS}
Rogers, F.~J., Swenson, F.~J., \& Iglesias, C.~A. 1996,
ApJ, 456, 902
\bibitem[Pandharipande \& Ravenhall(1989)]{Pandharipande89}
Pandharipande, V.~R., \& Ravenhall, D.~G. 1989,
Hot Nuclear Matter,
in Nuclear Matter and Heavy Ion Collisions,
NATO ADS Ser., vol.\ B205, ed. M.~Soyeur, H.~Flocard,
B.~Tamain, \& M.~Porneuf
 (Dordrecht: Reidel), 103
\bibitem[Stergioulas(2003)]{Sterg03}
Stergioulas, N. 2003, Living Rev.\ Relativity, 6, 3
(http://www.livingreviews.org/lrr-2003-3/)
\bibitem[Swesty(1996)]{Swesty96}
Swesty, F.~D. 1996, J.\ Comput.\ Phys., 127, 118
\end{thebibliography}
\end{document}